\documentclass[12pt,twoside]{article}
\usepackage{amssymb}
\usepackage{amsmath,amsthm}
\usepackage{latexsym}

\begin{document}
\begin{center}
{\Large {\bf Energy Current Correlations For \\[1ex]Weakly Anharmonic Lattices}}\footnote{Contribution to ICMP 15, Rio de Janeiro, August
2006}\bigskip\medskip\\
{\large{Herbert Spohn}}
\medskip\\
Zentrum Mathematik and Physik Department, TU M\"{u}nchen,\\
D - 85747 Garching, Boltzmannstr. 3, Germany, {\tt spohn@ma.tum.de}\\
\bigskip\bigskip\bigskip
\end{center}
\section{Introduction}\label{sec.1}
\setcounter{equation}{0}

A solid transports energy. Besides the mobile electrons, one
important mechanism for energy transport are the vibrations of the
crystal lattice. There is no difficulty in writing down the
appropriate lattice dynamics. To extract from it the thermal
conductivity remains a fairly untractable problem. The most
successful approach exploits that even rather close to the melting
temperature the typical deviations of the crystal atoms from their
equilibrium position are small as compared to the lattice constant.
This observation then leads to the phonon kinetic equation, which
goes back to the seminal paper by Peierls \cite{Pe29}. (For electron
transport a corresponding idea was put forward by Nordheim
\cite{No}.) Phonon kinetic theory flourished in the 50ies, an
excellent account of the 1960 status being the book by Ziman
\cite{Zi}. Of course, transport of heat and thermal conductivity
remain an important experimental research area, in particular since
novel materials become available and since more extreme properties
are in demand. On the other hand, if the very recent collection of
articles by Tritt \cite{Tr} is taken to be representative, it is
obvious that after 1960 hardly any new elements have been added to
the theory. The real innovation are fast and efficient molecular
dynamics algorithms. The currently available techniques allow the
simulation of $6\times 6\times 6$ periodized lattices with two atoms
per unit cell \cite{GK}.

According to the Green-Kubo formula the thermal conductivity is
determined through the time-integral over the energy current
correlation in thermal equilibrium. In my contribution I will
explain its structure for weakly anharmonic lattices. While I do not
add anything novel in substance, I believe that, with the post 1960
insights gained from the kinetic theory of rarified gases, the story
can be presented more concisely and systematically than done
usually. As a bonus, the mathematical physics issues left unresolved
will become more sharply in focus.

\section{Anharmonic lattice dynamics}\label{sec.2}
\setcounter{equation}{0}

Physically, one starts from a given crystal structure, which means
to specify the lattice and the number of atoms per unit cell. The
interaction potential is expanded in the displacements away from the
equilibrium positions. Then the first order term vanishes, because
one expands at a stationary point. The second order term is the
harmonic approximation and higher order terms are regarded as small
corrections. It is argued that for real crystals mostly the third
order term suffices unless there are special symmetries which make
it vanish and requires to go to fourth order. In this article, the
focus will be on the analysis of the linearized Boltzmann equation
and its relation to the energy current correlations. For this
purpose we take the liberty to employ a single band model for the
anharmonic lattice dynamics. There is no difficulty, in principle,
to add on extra features so to make the model more realistic.

We assume a simple hypercubic lattice $\mathbb{Z}^d$ with a single
atom per unit cell. Physically $d=3$, but we keep the general
dimension $d$ because of recent interest in chains, for which $d=1$.
A single band model corresponds to scalar atomic displacements.

Fourier transform will be convenient. Let
$\mathbb{T}^d=[-\frac{1}{2},\frac{1}{2}]^d$ be the first Brioullin
zone of the dual lattice. For $f:\mathbb{Z}^d\to\mathbb{R}$ its
Fourier transform, $\widehat{f}$, is defined by
\begin{equation}\label{2.1}
\widehat{f}(k)=\sum_{x\in\mathbb{Z}^d} e^{-i 2\pi k\cdot x}f_x\,.
\end{equation}
Here $k\in\mathbb{T}^d$ and $\widehat{f}(k)$ extends periodically to
a function on $\mathbb{R}^d$. The inverse Fourier transform is given
by
\begin{equation}\label{2.2}
f_x=\int_{\mathbb{T}^d}dk e^{i 2\pi k\cdot x} \widehat{f}(k)\,.
\end{equation}

For $x\in\mathbb{Z}^d$ the deviation away from $x$ is denoted by
$q_x\in\mathbb{R}$. The corresponding momentum is denoted by
$p_x\in\mathbb{R}$. We choose units such that the atomic mass equals
one. The harmonic approximation to the interaction potential reads
\begin{equation}\label{2.3}
U_\textrm{harm}(q)=\frac{1}{2}\sum_{x,y\in\mathbb{Z}^d}\alpha(x-y)q_x
q_y\,.
\end{equation}
The elastic constants $\alpha(x)$ satisfy
\begin{equation}\label{2.4}
\alpha(x)=\alpha(-x)\,,\quad |\alpha(x)|\leq \gamma_0
e^{-\gamma_1|x|}
\end{equation}
for suitable constants $\gamma_0,\gamma_1>0$. Mechanical stability
requires
\begin{equation}\label{2.5}
 \widehat{\alpha}(k)\geq 0\,.
\end{equation}
In addition, because of the invariance of the interaction between
crystals atoms under the translation $q_x\rightsquigarrow q_x+a$,
one imposes
\begin{equation}\label{2.6}
\sum_{x\in\mathbb{Z}^d}\alpha(x)=0\,,\quad
i.e.\;\;\widehat{\alpha}(0)=0\,.
\end{equation}
For an optical band, because of the internal structure of the unit
cell, the condition (\ref{2.6}) is not satisfied, which in the
framework of our model can be interpreted as adding to the physical
harmonic interaction satisfying (\ref{2.6}) a harmonic on-site
potential of the form
\begin{equation}\label{2.7}
U_\mathrm{site}(q)=\frac{1}{2}\omega^2_0\sum_{x\in\mathbb{Z}^d}q^2_x\,.
\end{equation}

The harmonic lattice dynamics is governed by the hamiltonian
\begin{equation}\label{2.8}
H_\mathrm{ha}=\frac{1}{2}\sum_{x\in\mathbb{Z}^d}(p^2_x+\omega^2_0
q^2_x)+ \frac{1}{2}\sum_{x,y\in\mathbb{Z}^d}\alpha(x-y)q_xq_y
\end{equation}
and has plane wave solutions with dispersion relation
\begin{equation}\label{2.9}
\omega(k)=(\omega^2_0+\widehat{\alpha}(k))^{1/2}\,.
\end{equation}
Clearly, $\omega(k)=\omega(-k)$ and $\omega(k)\geq\omega_0\geq0$. We
concatenate $q_x$ and $p_x$ into a single complex-valued field
$a(k)$ as
\begin{equation}\label{2.10}
a(k)=\frac{1}{\sqrt{2}}\big(\sqrt{\omega(k)}\widehat{q}(k)+
i\frac{1}{\sqrt{\omega(k)}}\widehat{p}(k)\big)
\end{equation}
with the inverse
\begin{equation}\label{2.11}
\widehat{q}(k)=\frac{1}{\sqrt{2}}\frac{1}{\sqrt{\omega(k)}}
\big(a(k)+a(-k)^\ast\big)\,,\quad
\widehat{p}(k)=\frac{i
}{\sqrt{2}}\sqrt{\omega(k)}
\big(-a(k)+a(-k)^\ast\big)\,.
\end{equation}
The $a$-field evolves as
\begin{equation}\label{2.12}
\frac{\partial}{\partial t}a(k,t)=-i\omega(k)a(k,t)\,.
\end{equation}

In nature lattice vibrations are quantized. In our model this is
easily implemented by promoting $a(k)^\ast$ and $a(k)$ to creation
and annihilation operators of a scalar Bose field. $a(k)^\ast$ is
the operator adjoint to $a(k)$ and the $a(k)$'s satisfy the
canonical commutation relations
\begin{equation}\label{2.13}
[a(k),a(k')^\ast]=\delta(k-k')\,,\;[a(k),a(k')]=0\,.
\end{equation}
The Heisenberg evolution for the $a$-field is still governed by
(\ref{2.12}).

Continuing the expansion scheme we add to $H_0$ the next order
terms. The simplest one would be a cubic on-site potential as
\begin{equation}\label{2.14}
V_3=\frac{1}{3}\sum_{x\in\mathbb{Z}^d} q^3_x\,,
\end{equation}
which in terms of the $a$-field reads
\begin{equation}\label{2.15}
V_3=\frac{1}{3}\int_{\mathbb{T}^{3d}}dk_1dk_2dk_3\delta(k_1+k_2+k_3)
\prod^3_{j=1}(2\omega(k_j))^{-1/2}\big(a(k_j)+a(-k_j)^\ast\big)\,.
\end{equation}
Correspondingly, at fourth order,
\begin{equation}\label{2.16}
V_4=\frac{1}{4}\sum_{x\in\mathbb{Z}^d}q^4_x
\end{equation}
which in terms of the $a$-field becomes
\begin{equation}\label{2.17}
V_4=\frac{1}{4}\int_{\mathbb{T}^{4d}}dk_1dk_2dk_3dk_4\delta(k_1+k_2+k_3+k_4)
\prod^4_{j=1}(2\omega(k_j))^{-1/2}\big(a(k_j)+a(-k_j)^\ast\big)\,.
\end{equation}

$H_\mathrm{ha}+\lambda V_3$ is not bounded from below. This can be
remedied by adding $\lambda^2 V_4$, for example, which would then
not contribute on the kinetic scale.

If the potential depends only on the displacement differences, then
the lowest order nonlinearity is
\begin{equation}\label{2.18}
V_{3\mathrm{di}}=\frac{1}{3}\sum_{x,y\in\mathbb{Z}^d}\alpha_3(x-y)(q_x-q_y)^3
\end{equation}
with $\alpha_3(-x)=-\alpha_3(x)$ and $|\alpha_3|$ exponentially
bounded. Switching to the $a$-field $V_{3\mathrm{di}}$ becomes
\begin{eqnarray}\label{2.19}
&&\hspace{-30pt}V_{3\mathrm{di}}=\frac{1}{3}\int_{\mathbb{T}^{4d}}
dk_1dk_2dk_3\delta(k_1+k_2+k_3)\nonumber\\
&&\hspace{24pt}\times\sum_{x\in\mathbb{Z}^d}\alpha_3(x)\prod^3_{j=1}
(2\omega(k_j))^{-1/2}\big(e^{i2\pi k_j\cdot x}-1\big)
\big(a(k_j)+a(-k_j)^\ast\big)\,.
\end{eqnarray}
In the kinetic limit the square of the vertex function determines
the collision rate. Thus, from the collision rate
\begin{equation}\label{2.20}
\prod^3_{j=1}(2\omega(k_j))^{-1}
\end{equation}
for the on-site $V_3$ the collision rate for $V_{3\mathrm{di}}$ is
obtained by the replacement
\begin{equation}\label{2.21}
\prod^3_{j=1}\big(2\omega(k_j)\big)^{-1}
\big|\sum_{x\in\mathbb{Z}^d}\alpha_3(x)\prod^3_{j=1}(e^{i2\pi
k_j\cdot x}-1)\big|^2\,.
\end{equation}
Because of such a simple substitution rule we continue to work with
$V_3$. The corresponding rule also applies to the switch from $V_4$
to $V_{4\mathrm{di}}$.

From other areas of mathematical physics one is accustomed to have a
given starting hamiltonian. In our context this means to specify the
elastic constants $\alpha(x)$, $\alpha_3(x)$, $\alpha_4(x)$. For
real crystals their determination requires a lot of experimental
(and also theoretical) efforts, as discussed in \cite{Tr}, see also
\cite{Wa,OS96,OS95} for a modeling of aluminium and silicon. It
would be thus of importance to have a stability result available,
which ensures that certain qualitative properties do not depend so
much on the specific choice of elastic constants.

\section{Energy current correlations}\label{sec.3}
\setcounter{equation}{0}

Let us consider the Hamiltonian
\begin{equation}\label{3.1}
H=H_\textrm{ha}+\lambda V_3+\lambda^2 V_4\,.
\end{equation}
The total energy current correlation function is computed in thermal
equilibrium at inverse temperature $\beta$. It is denoted by
$C_\lambda(t)$ and will be defined below. Since $\lambda\ll 1$, the
plan is to compute $C_\lambda(t)$ in the limit of $\lambda\to 0$.
The phonons then hardly interact and $C_\lambda(t)$ decays slowly on
the time scale $\lambda^{-2}$. Thus one expects that the limit
\begin{equation}\label{3.2}
\lim_{\lambda\to 0} C_\lambda(\lambda^{-2}t)=C_\textrm{kin}(t)
\end{equation}
exists and is determined by the phonon Boltzmann equation linearized
at equilibrium.

Let us first find the local energy current. Since $H$ is not local,
there is some arbitrariness involved in defining the local energy.
One conventional choice for the energy at site $x$ is to set
\begin{equation}\label{3.3}
H_x=\frac{1}{2}p^2_x+\frac{1}{2}\omega^2_0
q^2_x+\frac{1}{2}\sum_{y\in\mathbb{Z}^d}\alpha(x-y)q_xq_y+\frac{1}{3}\lambda
q^3_x+\frac{1}{4}\lambda^2q^4_x\,.
\end{equation}
In the Heisenberg picture $H_x$ becomes time-dependent. Writing
$dH_x(t)/dt$ as a divergence, the energy current can be identified
with
\begin{equation}\label{3.4}
J_x=\frac{1}{4}\sum_{y\in\mathbb{Z}^d}y\alpha(y)(-q_x
p_{x+y}+q_{x+y} p_x)
\end{equation}
which happens to be independent of $\lambda$. To verify (\ref{3.4}),
one chooses a large box $\Lambda$ with faces $\partial\Lambda$. The
energy inside $\Lambda$ is
\begin{equation}\label{3.5}
H_\Lambda(t)=\sum_{x\in\Lambda}H_x(t)
\end{equation}
and it satisfies
\begin{equation}\label{3.5a}
    \frac{d}{dt}H_\Lambda(t)=-\sum_{x\in\Lambda} n_x\cdot
    J_x(t)+\mathcal{O}(\partial\Lambda)\,,
\end{equation}
where $n_x$ is the outward normal to $\Lambda$ at
$x\in\partial\Lambda$. The errors come from the corners of $\Lambda$
and from the possibly infinite range of $\alpha$.

With this input the total energy current correlation is defined by
\begin{equation}\label{3.6}
\ell\cdot C_\lambda(t)\ell=\sum_{x\in\mathbb{Z}^d}\langle (\ell\cdot
J_0(t))(\ell\cdot J_x(0))\rangle_\beta\,,
\end{equation}
where $\ell\in\mathbb{R}^d$, $J_x\in\mathbb{R}^d$, ``$\,\cdot\,$''
is the scalar product in $\mathbb{R}^d$, and $C_\lambda(t)$ is a
$d\times d$ matrix. $\langle\cdot\rangle_\beta$ refers to the
thermal average with respect to $Z^{-1}e^{-\beta H}$. By
time-stationary and time-reversal, $C_\lambda(t)$ is symmetric and
it suffices to consider its numerical range. At the expense of an
error of order $\lambda$, we may replace in $e^{-\beta H}$ the full
Hamiltonian $H$ by the harmonic approximation
\begin{equation}\label{3.7}
H_\mathrm{ha}=\int_{\mathbb{T}^d}dk\omega(k)a(k)^\ast a(k)\,.
\end{equation}
For the total current one finds
\begin{equation}\label{3.8}
J=\sum_{x\in\mathbb{Z}^d}J_x=\frac{1}{2\pi}
\int_{\mathbb{T}^d}dk(\nabla\omega(k))\omega(k)a(k)^\ast a(k)\,,
\end{equation}
where it is used that
$\nabla\widehat{\alpha}=2\omega(\nabla\omega)$. Since
$[H_\textrm{ha},J]=0$, $\sum_{x\in\mathbb{Z}^d}\ell\cdot J_x$ can be
lifted to the exponent. Thus we define the new average
$\langle\cdot\rangle_{\beta,\tau}$ with respect to the state
$Z^{-1}\exp[-\beta H_\textrm{ha}+\tau\ell\cdot J]$. Then
\begin{equation}\label{3.9}
\ell\cdot C_\lambda(t)\ell=\lim_{\tau\to
0}\frac{1}{\tau}\langle\ell\cdot
J_0(t)\rangle_{\beta,\tau}+\mathcal{O}(\lambda)\,.
\end{equation}
The anharmonicity now resides only in the dynamics.

The limit $\lambda\to 0$ on the right hand side in (\ref{3.9}) is
discussed in \cite{Sp06}. The initial state is spatially homogeneous
and determines the Wigner function $W_{\beta,\tau}(k)$ through
\begin{equation}\label{3.10}
\langle a(k')^\ast
a(k)\rangle_{\beta,\tau}=\delta(k-k')W_{\beta,\tau}(k)
\end{equation}
with
\begin{equation}\label{3.11}
W_{\beta,\tau}(k)=\Big(\exp\big[\beta\omega(k)-
\tau\big(\ell\cdot\nabla\omega(k)\big)\omega(k)\big]-1\Big)^{-1}\,.
\end{equation}
On the kinetic time scale, $\lambda^{-2}t$, the Wigner function
$W_{\beta,\tau}$ evolves to $W_\tau(t)$ which is determined as the
solution of the spatially homogeneous Boltzmann equation. For our
model, i.e.~for the anharmonic on-site potential $V_3$, it reads
\begin{equation}\label{3.11a}
\frac{\partial}{\partial t}W(t)=\mathcal{C}\big(W(t)\big)
\end{equation}
with the collision operator
\begin{eqnarray}\label{3.12}
&&\hspace{-50pt}\mathcal{C}(W)_1=
\frac{\pi}{2}\int_{\mathbb{T}^{2d}} d k_2 d k_3
(\omega_1\omega_2\omega_3)^{-1}\nonumber\\
&&\hspace{16pt}\times\big\{2\delta (\omega_1+\omega_2-\omega_3)
\delta(k_1+k_2-k_3)\big(\tilde{W}_1 \tilde{W}_2 W_3
- W_1 W_2 \tilde{W}_3\big)\nonumber\\
&&\hspace{30pt}+\; \delta (\omega_1-
\omega_2-\omega_3)\delta(k_1-k_2-k_3)\big(\tilde{W}_1 W_2 W_3
-W_1\tilde{W}_2\tilde{W}_3\big)\big\}\,.
\end{eqnarray}
Here we use the shorthands $W_j=W(k_j)$, $\omega_j=\omega(k_j)$,
$j=1,2,3$, and $\tilde{W}(k)=1+W(k)$. Using (\ref{3.8}), the average
in (\ref{3.9}) becomes then
\begin{equation}\label{3.13}
\langle\ell\cdot J_0(\lambda^{-2}t)\rangle_{\beta,\tau}=
\frac{1}{2\pi}\int_{\mathbb{T}^d}dk\big(\ell\cdot\nabla\omega(k)\big)
\omega(k)W_\tau(k,t)+\mathcal{O}(\lambda)\,.
\end{equation}
The next task is to take the limit $\tau\to 0$ in (\ref{3.9}). One
has $W_{\beta,0}(k)=(e^{\beta\omega(k)}-1)^{-1}$ which is a
stationary solution of (\ref{3.11a}). Thus the limit $\tau\to 0$
amounts to linearize (\ref{3.12}) at the equilibrium Wigner function
\begin{equation}\label{3.14}
W_\beta(k)=(e^{\beta\omega(k)}-1)^{-1}\,,
\end{equation}
to say
\begin{equation}\label{3.15}
W_{\beta,\tau}=W_\beta+\tau
W_\beta\tilde{W}_\beta(\ell\cdot\nabla\omega)
\omega+\mathcal{O}(\tau^2)\,.
\end{equation}
Note that $\int dk(\nabla\omega)\omega W_\beta=0$. As suggested by
(\ref{3.15}), with a significance which will become more convincing
in the context of the Gaussian fluctuation theory, see Section
\ref{sec.5}, the natural linearization of $\mathcal{C}$ is
\begin{equation}\label{3.16}
\mathcal{C}(W_\beta+\delta W_\beta\tilde{W}_\beta f)=-\delta L
f+\mathcal{O}(\delta^2)\,.
\end{equation}
From (\ref{3.12}) one deduces
\begin{eqnarray}\label{3.17}
&&\hspace{-46pt}(L f)_1= \frac{\pi}{2}\int_{\mathbb{T}^{2d}} d k_2 d
k_3(\omega_1\omega_2\omega_3)^{-1}\nonumber\\
&&\hspace{16pt} \times\big(2\delta
(\omega_1+\omega_2-\omega_3)\delta(k_1+k_2-k_3)\tilde{W}_{\beta 1}
\tilde{W}_{\beta 2} W_{\beta 3}(f_1+f_2-f_3)\nonumber\\
&&\hspace{30pt}+\delta
(\omega_1-\omega_2-\omega_3)\delta(k_1-k_2-k_3)\tilde{W}_{\beta 1}
W_{\beta 2} W_{\beta 3}(f_1-f_2-f_3)\big)\,.
\end{eqnarray}
Properties of $L$ will be discussed in the subsequent section.

Let  $A$ be the linear operator obtained from flat linearization as
\begin{equation}\label{3.18}
\mathcal{C}(W_\beta+\delta f)=\delta A f+\mathcal{O}(\delta^2)\,.
\end{equation}
Clearly $A(W_\beta\tilde{W}_\beta f)=-L f$. Combining (\ref{3.9})
and (\ref{3.13}) we finally conclude
\begin{equation}\label{3.19}
\lim_{\lambda\to 0}\ell\cdot C_\lambda(\lambda^{-2}t)\ell=\ell\cdot
C_\textrm{kin}(t)\ell
\end{equation}
with
\begin{equation}\label{3.20}
\ell\cdot C_\textrm{kin}(t)\ell=\langle
(2\pi)^{-1}(\ell\cdot\nabla\omega)\omega\,,\;e^{-A|t|} W_\beta
\tilde{W}_\beta (2\pi)^{-1}(\ell\cdot\nabla\omega)\omega\rangle\,,
\end{equation}
where $\langle\cdot,\cdot\rangle$ is the inner product in
$L^2(\mathbb{T}^d,dk)$.

For future use it will be convenient to write $C_\textrm{kin}(t)$ in
a more symmetric form. Expanding the exponential one notes that
\begin{eqnarray}\label{3.21}
&&\hspace{-45pt}\ell\cdot C_\textrm{kin}(t)\ell=
\langle(2\pi)^{-1}(\ell\cdot\nabla\omega)\omega
(W_\beta\tilde{W}_\beta)^{1/2}\,,\nonumber\\[1ex]
&&\hspace{-8pt}\exp\big[-(W_\beta\tilde{W}_\beta)^{-(1/2)}
L(W_\beta\tilde{W}_\beta)^{-(1/2)}|t|\big](W_\beta\tilde{W}_\beta)^{1/2}(2\pi)^{-1}
(\ell\cdot\nabla\omega)\omega\rangle\,.
\end{eqnarray}
As will be shown, $L=L^\ast$, i.e.~$L$ is a symmetric operator in
$L^2(\mathbb{T}^d,dk)$. Therefore $C_\textrm{kin}(t)$ is a positive
symmetric $d\times d$ matrix.

In the kinetic limit the thermal conductivity is given through
\begin{eqnarray}\label{3.22}
&&\hspace{-20pt}\ell\kappa_\textrm{kin}\cdot\ell=
\beta^2\int^\infty_0 dt \ell\cdot
C_\textrm{kin}(t)\ell\nonumber\\
&&\hspace{21pt}=(2\pi)^{-2}\beta^2\langle(\ell\cdot\nabla\omega)\omega
W_\beta\tilde{W}_\beta\,,\; L^{-1}(\ell\cdot\nabla\omega)\omega
W_\beta\tilde{W}_\beta\rangle\,.
\end{eqnarray}
Reversing our argument, and assuming uniformity in $t$ for the limit
$\lambda\to 0$, one infers that the true thermal conductivity,
$\kappa(\lambda)$, of the anharmonic model behaves as
\begin{equation}\label{3.23}
\kappa(\lambda)\cong \lambda^{-2}\kappa_\textrm{kin}
\end{equation}
for small $\lambda$.

In the classical limit $[a(k),a(k')^\ast]=0$, i.e., $W=\tilde{W}$.
In the definition of $L$ one has thus to replace
\begin{equation}\label{3.24}
W_\beta\,,\;\tilde{W}_\beta\quad \textrm{by}\quad
W^{\mathrm{cl}}_\beta (k)=\frac{1}{\beta\omega(k)}\,.
\end{equation}

We presented the argument for a cubic on-site potential. But,
clearly, the result holds also for other small anharmonicities. Only
the collision operator, and its linearization $L$, would have to be
modified.

\section{The linearized collision operator}\label{sec.4}
\setcounter{equation}{0}

If one accepts the argument leading to (\ref{3.19}), the remaining
task is to study the spectral properties of the linearized collision
operator, from which the time decay of $C_\mathrm{kin}(t)$ can be
infered. While this looks like a conventional mathematical physics
problem, the difficulty comes from the energy-momentum constraint.
Only in a few special cases there is an explicit solution. Otherwise
one has to work with the implicit definition. In fact, there can be
no solution at all, in which case $L=0$, or several solutions, in
which case one has to sum over all collision branches.\smallskip\\
(i) \textit{quadratic form}. For three phonon processes, on-site
potential $V_3$, the quadratic form of the linearized collision
operator $L=L_3$ is given by
\begin{eqnarray}\label{4.1}
&&\hspace{-28pt}\langle g, L_3 f\rangle=\frac{\pi}{2}
\int_{\mathbb{T}^{3d}} d k_1 d k_2 d
k_3(\omega_1\omega_2\omega_3)^{-1} \delta
(\omega_1+\omega_2-\omega_3)\delta(k_1+k_2-k_3)\nonumber\\
&&\hspace{50pt}\times W_{\beta 1} W_{\beta 2} \tilde{W}_{\beta
3}(g_1+g_2-g_3)(f_1+f_2-f_3)\,.
\end{eqnarray}
Correspondingly for the on-site potential $V_4$ one has
\begin{eqnarray}\label{4.2}
&&\hspace{-20pt}\langle g, L_4
f\rangle\nonumber\\
&&\hspace{-10pt}
=\frac{3\pi}{4}\cdot\frac{3}{4} \int d k_1 d k_2 d k_3 d k_4
(\omega_1\omega_2\omega_3\omega_4)^{-1} \delta
(\omega_1+\omega_2-\omega_3-\omega_4)
\delta(k_1+k_2-k_3-k_4)
\nonumber\\
&&\hspace{38pt}\times W_{\beta 1} W_{\beta 2} \tilde{W}_{\beta
3}\tilde{W}_{\beta 4}(g_1+g_2-g_3-g_4)(f_1+f_2-f_3-f_4)\nonumber\\
&&\hspace{4pt}+ \frac{3\pi}{4}\int d k_1 d k_2 d k_3 d k_4
(\omega_1\omega_2\omega_3\omega_4)^{-1} \delta
(\omega_1+\omega_2+\omega_3-\omega_4)\delta(k_1+k_2+k_3-k_4)\nonumber\\
&&\hspace{38pt}\times W_{\beta 1} W_{\beta 2} W_{\beta
3}\tilde{W}_{\beta 4}(g_1+g_2+g_3-g_4)(f_1+f_2+f_3-f_4)\nonumber\\
&&\hspace{-10pt}=\langle f,L_{4\mathrm{p}}f\rangle+\langle
f,L_{4\mathrm{t}}f\rangle\,.
\end{eqnarray}
$L_{4\mathrm{p}}$ corresponds to the collision of a pair of phonons
and $L_{4\mathrm{t}}$ to a merger of three phonons into a single
one, and its time reversal. The quadratic forms for
$L_3,L_{4\mathrm{p}},L_{4\mathrm{t}}$ (for notational simplicity from now on commonly
denoted by $L$) are somewhat formal. Firstly, if $\omega(0)=0$ and
$\omega(k)>0$ otherwise, the smooth functions $g,f$ have to vanish
at $k=0$. More seriously, the proper definition of the
$\delta$-function requires to study more carefully the solutions to
the energy constraint
\begin{equation}\label{4.2a}
\omega(k_1)+\omega(k_2)=\omega(k_1+k_2)\,,
\end{equation}
say in the case of $L_3$. For the purpose of our exposition, let us
simply assume that the quadratic form defines $L$ as a
self-adjoint operator. Clearly, $L\geq 0$ since $\langle f,L
f\rangle\geq 0$. As $(\ell\cdot\nabla\omega)\omega$ is bounded, one
has
\begin{equation}\label{4.3}
\ell\cdot C_\mathrm{kin}(t)\ell\leq
(2\pi)^{-2}\langle(\ell\cdot\nabla\omega)\omega, (\ell\cdot
\nabla\omega)\omega\rangle\,.\smallskip
\end{equation}
(ii) \textit{zero subspace}. To establish that
$\lim_{t\to\infty}C_\mathrm{kin}(t)=0$,
$(\ell\cdot\nabla\omega)\omega$ has to be orthogonal to the zero
subspace of $L$. There seems to be no cheap argument and one has to
study the solutions to
\begin{equation}\label{4.4}
L f=0\,.
\end{equation}
From (\ref{4.1}), (\ref{4.2}) it follows that $f$ has to be a
collisional invariant, see \cite{Sp06,Sp06/2} for the definition.
Considering only the first summand of (\ref{4.2}), there is a
general argument \cite{Sp06/2}, that the solutions to $\langle f,
L_{4\mathrm{p}} f\rangle=0$ are spanned by 1, $\omega$. Note that
$\langle 1,(\ell\cdot\nabla\omega)\omega\rangle=0$,
$\langle\omega,(\ell\cdot \nabla\omega)\omega\rangle=0$. The
constant function results from phonon number conservation in a pair
collision. This conservation law will be destroyed by adding a
little bit of either three-phonon, $L_3$, or the second term of the
four-phonon processes, $L_{4\mathrm{t}}$. The zero subspace is then
one-dimensional and spanned by $\omega$ only. For $L_3$ of
(\ref{4.1}), the classification of the collisional invariants is an
open problem.
\smallskip\\
(iii) \textit{spectral gap}. If $L$ has a spectral gap, the energy
current correlation decays exponentially. If in addition
$(\ell\cdot\nabla\omega)\omega$ is orthogonal to the  zero subspace
of $L$, then the conductivity, as the time-integral over
$C_\mathrm{kin}(t)$, is finite (and non-zero). In particular
(\ref{3.23}) holds.

$L$ is a sum of a multiplication operator and an integral operator,
\begin{equation}\label{4.5}
L=V+I\,,\quad V f(k)=V(k)f(k)\,,\quad I f(k)=\int_{\mathbb{T}^d} dk'
I(k,k')f(k')\,,
\end{equation}
where, say in the case of $L_3$,
\begin{eqnarray}\label{4.6}
&&\hspace{-48pt}V(k)=\frac{\pi}{2}
W_\beta(k)\omega(k)^{-1}\int_{\mathbb{T}^d} d k_1
\big(\omega(k_1)\omega(k+k_1)\big)^{-1}\nonumber\\
&&\hspace{-4pt}\times \Big (2\delta
(\omega(k)+\omega(k_1)-\omega(k+k_1))W_\beta(k_1)\tilde{W}_\beta(k+k_1)
\nonumber\\
&&\hspace{8pt}+\; \delta (\omega(k)-
\omega(k_1)-\omega(k+k_1))\tilde{W}_\beta(k_1)\tilde{W}_\beta(k+k_1)\Big)\,.
\end{eqnarray}
The integral kernel $I(k,k')$ is implicitly defined. It has no
definite sign and tends to be divergent on lower-dimensional
submanifolds of $\mathbb{T}^d\times\mathbb{T}^d$. It would be useful
to know under what conditions the integral operator $I$ is compact.

In the very common relaxation time approximation, $I$ is simply
dropped and one sets in approximation
\begin{equation}\label{4.8}
\ell\cdot C_\textrm{kin}(t)\ell=(2\pi)^{-2}
\langle(\ell\cdot\nabla\omega)
\omega,e^{-|t|/\tau}(\ell\cdot\nabla\omega)\omega\rangle
\end{equation}
 with the relaxation time
\begin{equation}\label{4.7}
\tau(k)=W_\beta(k)\tilde{W}_\beta(k) V(k)^{-1}\,,
\end{equation}
see (3.23). \smallskip\\
(iv) \textit{FPU chains}. The Fermi-Pasta-Ulam chain is the special
case $d=1$ with nearest neighbor coupling and no quantization. For a
harmonic on-site potential the dispersion relation is
$\omega(k)=\big(\omega^2_0+1-\cos(2\pi k)\big)^{1/2}$,
$k\in\mathbb{T}$. Although $d=1$, the conservation laws of energy
and momentum allow for non-degenerate pair collision and
$L_{4\mathrm{p}}\neq 0$, while $L_{4\mathrm{t}}=0$ \cite{Pe,ALS,LS}.
There are fairly explicit formuli for the potential $V$ and the
integral kernel $I$ \cite{Lu}. For $\omega_0>0$ and a quartic
on-site potential $V_4$, the linearized collision operator has a gap
and the zero subspace is two-dimensional. The gap seems to close as
$\omega_0\to 0$. On the basis of numerical simulations, the
conductivity should be finite even for $\omega_0=0$ \cite{ALS}. The
FPU-$\beta$ chain has the nonlinearity $V_{\mathrm{di} 4}$. $V$ and
$I$ has been computed by Pereverzev \cite{Pe}. He uses the
relaxation time approximation and finds that $C_\mathrm{kin}(t)\cong
t^{-3/5}$ for large $t$. Using a resolvent expansion, in  \cite{LS} we prove corresponding sharp bounds and thereby confirm the relaxation time approximation in this particular case. For a finite chain of length $N$ with
thermal reservoirs at both ends, the energy transport is then anomalous and the thermal conductivity diverges as $N^{2/5}$,
which seems to be in agreement with molecular dynamics. For a more
detailed discussion we refer to \cite{LLP}, Section 6.

\section{Gaussian fluctuation theory}\label{sec.5}
\setcounter{equation}{0}

Energy transport can be viewed in the more general context of
time-dependent Gaussian fluctuation theory close to thermal
equilibrium. For low density gases this link is reviewed in
\cite{Sp83} with further examples discussed in \cite{Sp91}. The
purpose of this section is to explain how phonon kinetic theory
makes no exception. In \cite{Sp83,Sp91} spatial variation is
included. Since our exposition deals only with the spatially
homogeneous system, we stick to such a set-up also for the
fluctuation theory.

Physically, one considers time-dependent fluctuations in equilibrium
for the number of phonons with wave number $k$. Technically one has
to sum over phonons in a small volume element in $k$-space. To be
more precise we partition the tours $\mathbb{T}=[-1/2,1/2]$ by a
grid with spacing $\varepsilon$ and denote it by
$\mathbb{T}_\varepsilon$. $(\mathbb{T}_\varepsilon)^d$ corresponds
to the crystal volume $[1,\ldots,l]^d\subset\mathbb{Z}^d$ with
periodic boundary conditions, $l=1/\varepsilon$. Let
$f:\mathbb{T}^d\to\mathbb{R}$ be a smooth test function. Then the
fluctuation field, indexed by $f$ and $t$, is defined through
\begin{equation}\label{5.1}
\xi^\varepsilon(f,t)=\varepsilon^{d/2}
\sum_{k\in(\mathbb{T}_\varepsilon)^d}f(k)\big(a^\varepsilon(k,t)^\ast
a^\varepsilon(k,t)- \langle a^\varepsilon(k)^\ast
a^\varepsilon(k)\rangle_\beta\big)\,.
\end{equation}
$a^\varepsilon(k,t)$ depends on $\varepsilon$ through the finite
crystal volume $\varepsilon^{-d}$, through setting
$\lambda^2=\varepsilon$, and through the rescaled time
$\varepsilon^{-1}t$ in microscopic units. The claim is that, in
distribution, the limit
\begin{equation}\label{5.2}
\lim_{\varepsilon\to 0} \xi^\varepsilon(f,t)=\xi_t(f)
\end{equation}
exists and that the limit random field $\xi_t(f)$ is classical. In
fact, the limit field should be jointly Gaussian and governed by the
linear Langevin equation
\begin{equation}\label{5.3}
\frac{\partial}{\partial t}\xi_t(k)=A \xi_t(k)+B\eta_t(k)\,,
\end{equation}
where $\xi_t(f)=\int_{\mathbb{T}^d}dkf(k)\xi_t(k)$. $A$ is the
generator from the linearized Boltzmann equation, compare with
(\ref{3.18}), and $\eta_t$ is normalized Gaussian white noise with
\begin{equation}\label{5.4}
\mathbb{E}(\eta_t(k)\eta_{t'}(k'))=\delta(t-t')\delta(k-k')\,.
\end{equation}
The linear operator $B$ controls the strength and correlations for
the noise input to the various $k$-modes.

The main observation of the fluctuation theory is the relationship
between $A$ and $B$ through the equal-time equilibrium fluctuations.
We set, as a linear operator,
\begin{equation}\label{A.5}
\langle g\,,\;C f\rangle= \lim_{\varepsilon\to 0}\langle
\xi^\varepsilon(g,0)\xi^\varepsilon (f,0)\rangle_\beta\,.
\end{equation}
Using that
\begin{eqnarray}\label{5.6}
&&\hspace{-20pt} \langle
a^\ast(k_1)a(k_2)a^\ast(k_3)a(k_4)\rangle_\beta - \langle
a^\ast(k_1)a(k_2)\rangle_\beta\langle a(k_3)^\ast
a(k_4)\rangle_\beta
\nonumber\\
&&\hspace{30pt}=
\delta(k_1-k_4)\delta(k_2-k_3)W_\beta(k_1)\tilde{W}_\beta(k_2)\,,
\end{eqnarray}
one obtains
\begin{equation}\label{5.7}
\langle g\,,\;C f\rangle=\int_{\mathbb{T}^d}dk
g(k)W_\beta(k)\tilde{W}_\beta(k) f(k)\,,
\end{equation}
in other words $C$ is the operator of multiplication by
$W_\beta\tilde{W}_\beta$. The fluctuation-dissipation relation takes
then the form
\begin{equation}\label{5.8}
AC+CA^\ast=-BB^\ast\,.
\end{equation}
Since $AC=L=L^\ast=CA^\ast$, one concludes that the noise strength
is
\begin{equation}\label{5.9}
BB^\ast=2L\,.
\end{equation}

A posteriori this identity explains also the at first sight
unexpected linearization in (\ref{3.16}). Only then the linearized
operator is symmetric, as is obvious from (\ref{5.9}).

Solving (\ref{5.3}), the covariance of the stationary fluctuation
field is given by
\begin{equation}\label{5.10}
\langle \xi_t(g)\xi_0(f)\rangle= \langle g\,,\;e^{A|t|}C f\rangle\,,
\end{equation}
in agreement with the special case
$f=g=(2\pi)^{-1}(\ell\cdot\nabla\omega)\omega$ of interest in
Section \ref{sec.3}.

\end{document}